\documentclass[proceedings]{JHEP3}

\PrHEP{PrHEP hep2001}                  
\conference{International Europhysics Conference on HEP}          

\usepackage{amssymb}
\usepackage{epsfig}
\def\unit{\leavevmode\hbox{\small1\kern-3.6pt\normalsize1}}
\def \be{\begin{equation}}
\def \ee{\end{equation}}
\def \bea{\begin{eqnarray}}
\def \eea{\end{eqnarray}}
\def \ben{\begin{enumerate}}
\def \een{\end{enumerate}}
\def \bit{\begin{itemize}}
\def \eit{\end{itemize}}
\def \av#1{\left\langle #1\right\rangle}
\def \eff{\mathrm{eff}}
\def \vckm{V_{\mathrm{CKM}}}
\def \subfermion{\mathrm{fermion}}
\def \subhiggs{\mathrm{Higgs}}
\def \sb{\mathrm{SB}}
\def \susy{\mathrm{SUSY}}
\def \GeV{{\mathrm{GeV}}}

\def \TeV{{\mathrm{TeV}}}
\def \Im{{\mathrm{Im}}\,}

\def \bm{\boldmath}

\def \cl#1{{#1\%\ \mathrm{C.L.}}}

\def \cp{\mathrm{CP}}

\def \diag{{\mathrm{diag}}}

\def \ea{{\it et al.}}
\def \eq#1{Eq.~(\ref{#1})}

\def \heff{H_{\mathrm{eff}}}
\def \hc{\mathrm{H.c.}}
\def \nnu{\nonumber}

\def \rf{Ref.~\cite}

\def \b{\beta}
\def \f{\phi}
\def \D{\Delta}
\def \g{\gamma}

\def \d{\delta}
\def \epsi{\epsilon}

\def \l{\lambda}

\def \m{\mu}

\def\21{$SU(2) \ot U(1)$} 
\def\ot{\otimes}

\def\vev#1{\left\langle #1\right\rangle}

\def\bold#1{\setbox0=\hbox{$#1$} 
     \kern-.025em\copy0\kern-\wd0 
     \kern.05em\copy0\kern-\wd0 
     \kern-.025em\raise.0433em\box0 }
\def \chargino{\tilde{\chi}^{\pm}}

\def \gluino{\tilde{g}}
\def \higgs{H^{\pm}}
\def \neutralino{\tilde{\chi}^0}

\def \wino{\tilde{W}}

\def \higgsino{\tilde{H}}
\title{Spontaneous \bm$\cp$ Violation in the Next-to-Minimal
Supersymmetric Standard Model}

\author{\speaker{Ana M. Teixeira}                       
        \thanks{A.M.T. acknowledges support by  
`Funda\c c\~ao para a Ci\^encia e Tecnologia' under grant PRAXIS XXI 
BD/11030/97.} $^a$, \\                                  
      $^a$  Centro de F\'\i sica das Interac\c{c}\~{o}es Fundamentais (CFIF),
Departamento de F\'{\i}sica,  Instituto Superior T\'ecnico,  
Av. Rovisco Pais,  1049-001 Lisboa, Portugal\\                     
        E-mail: \email{ana@cfif.ist.utl.pt}}                       
\author{G.C. Branco$^a$, F. Kr\"uger$^b$ and J.C. Rom\~ao$^a$ \\
$^b$Physik Department, Technische Universit\"at M\"unchen,
D-85748 Garching, Germany\\                                        
E-mail:\email{gbranco@cfif.ist.utl.pt}, \email{fkrueger@ph.tum.de}, 
\email{fromao@alfa.ist.utl.pt}}
\abstract{We re-examine 
spontaneous $\cp$ violation (SCPV) at the tree level in the context of 
the next-to-minimal supersymmetric standard model (NMSSM) 
with two Higgs doublets and a gauge singlet field. 
We analyse the most general Higgs potential without a discrete 
$Z_3$ symmetry, and derive an upper bound on the mass of the lightest
neutral Higgs boson.
We estimate $\epsi_K$ by applying the mass insertion approximation, 
finding that in order to account for the for the observed 
$\cp$ violation in the neutral kaon sector 
a non-trivial flavour structure in the soft-breaking $A$ terms is required 
and that 
the upper bound on the lightest Higgs-boson mass becomes stronger.
We also discuss the implications of electric dipole moments 
of the electron and the neutron in SUSY models with SCPV. }
\begin{document}
\section{Introduction}\label{introduction}
As first proposed by T.D.Lee \cite{TDlee}, 
an alternative scenario for the breaking of $\cp$ is to assume that it is
a symmetry of the Lagrangian which is only spontaneously broken
by the vacuum. In Ref.~\cite{SCPV} we study the spontaneous 
breaking of $\cp$  at the tree level within the context of
supersymmetry (SUSY). 
We consider a simple extension of the MSSM with
one gauge singlet field ($N$) besides the two Higgs
doublets ($H_{1,2}$), the so-called next-to-minimal supersymmetric 
standard model (NMSSM).
In this class of models CP
violation is caused by the phases 
associated with the vacuum expectation values of the Higgs fields,
thus the reality of the 
CKM matrix is automatic and not an \emph{ad hoc} assumption.
The purpose of our work is to ask if 
one can achieve spontaneous breaking of $\cp$ whilst generating the
observed amount of $\epsi_K$ and having Higgs-boson masses
that are consistent with experimental data.
%
%
\section{The Higgs potential}\label{higgspotential}
We consider the most general form of the superpotential given by 
$W~=~W_{\subfermion}~+~W_{\subhiggs}$.
In addition to the usual MSSM terms, one finds new
contributions in $W_{\subhiggs}$, given by:
\be\label{pot:eq:WHiggs}
W_{\subhiggs}= -\lambda \widehat N 
\widehat H_1\widehat H_2-\frac{k}{3}\, {\widehat N}^3 -r \widehat N
-\mu  \widehat H_1\widehat H_2,
\ee
where $\widehat N$ is a singlet superfield.
Decomposing the SUSY soft-breaking terms as 
${\mathcal L}_{\sb}={\mathcal L}_{\sb}^{\subfermion}+
{\mathcal L}_{\sb}^{\subhiggs}$,
additional soft terms will appear in ${\mathcal L}_{\sb}^{\subhiggs}$ 
\be\label{pot:eq:softH}
-{\mathcal L}_{\sb}^{\subhiggs}=
m_{H_i}^2 H^{a*}_i H^a_i + m_N^2 N^* N\nnu
-\left(B\mu \varepsilon_{ab}H_1^a H_2^b +A_{\lambda} N \varepsilon_{ab} 
H_1^a H_2^b +\frac{A_k}{3}N^3 +A_r N + \hc \right).
\ee
In this analysis, we do not require the superpotential
to be invariant under a discrete $Z_3$ symmetry (which would imply 
$\mu=r=0$), nor do we relate the soft SUSY-breaking parameters  
to some common unification scale, but rather take them as arbitrary 
at the electroweak scale.
Throughout we shall assume that the tree-level 
potential is $\cp$ conserving and take all parameters real, but allow 
complex vacuum expectation values (VEV's) for the neutral Higgs
fields which  emerge after spontaneous symmetry breaking: 
$\vev{H_i^0}=v_i e^{i \theta_i}/\sqrt{2}$ and 
$\vev{N}=v_3 e^{i \theta_3}/\sqrt{2}$.
After deriving the $\cp$-invariant neutral scalar potential, it turns
out that only the following phase combinations are relevant:
$\phi_D=\theta_1+\theta_2,\; \phi_N=\theta_3$.\\
\vspace*{-1mm}
\EPSFIGURE{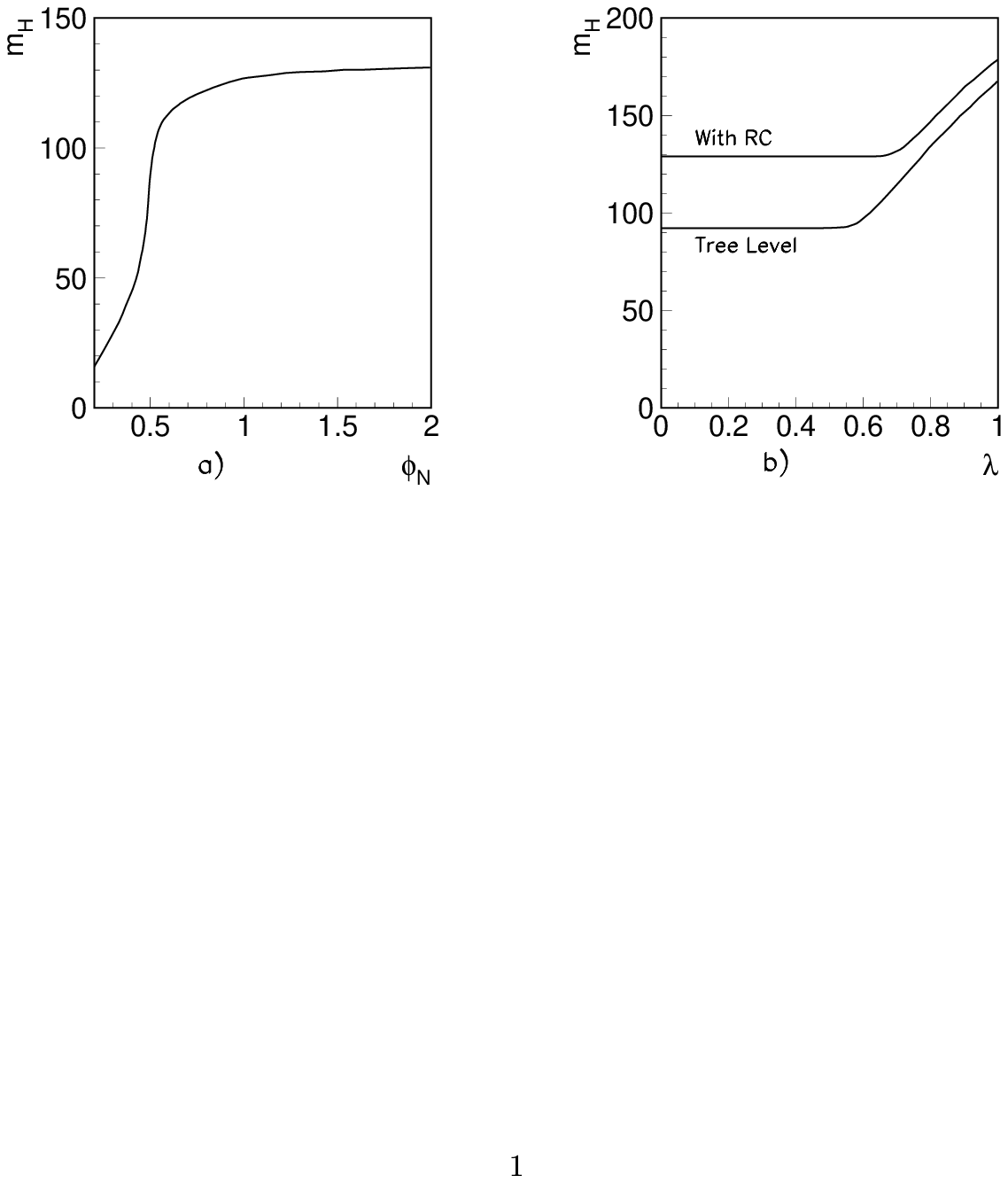,height=2.3in}
{Maximum value of the lightest Higgs-boson mass (in GeV) as a
function of the $\cp$-violating phase $\phi_N$ (in radians) 
(a), and as a function of the singlet coupling $\lambda$ 
at the tree level and after 
including radiative corrections (at one-loop level) for 
$M_{\susy}=1\,\TeV$ (b).\label{scpv_plot_mhtheta}}
\vspace*{-1mm}
We find that an acceptable mass spectrum can be easily obtained, with  
the exact values depending on the set of parameters we choose. As it
can be seen in Figure~\ref{scpv_plot_mhtheta}~a), the large singlet phase 
solution is favoured. The maximal possible value of the Higgs-boson
mass 
can differ from that of the MSSM for the case of large values of 
the coupling constant $\lambda$ as depicted in  
Figure~\ref{scpv_plot_mhtheta}~b).
For low values of $\lambda$, corrections to the tree level Higgs-boson
mass are significant and depend mainly on the SUSY scale that we take for
the squarks, with $max(m_{H^0})$ ranging from $105$ to $130 \GeV$, as the
typical SUSY scale varies from $300$ to $1000 \GeV$.
Finnaly, we point out that the SM and MSSM Higgs boson mass limits
obtained at LEP do not necessarily apply to the NMSSM (see, e.g., 
\rf{Higgs:mass}) since due to some 
singlet admixture the lightest neutral
Higgs boson 
may have a reduced coupling to the $Z^0$ \cite{model:nmssm:I}                  
and thus even escape detection.
%
%
\section{Brief overview of the model}\label{model}
In the scenario we are considering, CP invariance is imposed on the
lagrangian, and hence all couplings are real. Moreover, the $\vckm$ is
naturally real \cite{SCPV}.
Even so, the phases associated with the VEV's, $\f_{D}$ and $\f_{N}$,  
appear in the scalar quark, gaugino and Higgsino mass matrices, 
as well as in some of the vertices.\\ 
In the squark sector, working
in the `super-CKM' basis, we find complex contributions to the $LR$
submatrices of the up and down squark squared masses.
\bea\label{mass:squark:detail}
M^2_{\tilde{U}_{LR}} &=&M^{2\dagger}_{\tilde{U}_{RL}}=
V_L^U Y_U^*V_R^{U\dagger}\frac{v_2}{\sqrt{2}} 
- \m_{\eff} \cot\b e^{i \f_D} m_U^{\diag}\;; \; (U\rightarrow D),
\eea
where $Y_U^{ij}\equiv A_U^{ij} h_q^{ij}, \;
(\mathrm{no\ sum\ over\ }i,j), \;
\m_{\eff}\equiv \m +\l  \frac{v_3}{\sqrt{2}}e^{i\f_N}.$\\
In the chargino sector (defining $m_{ \tilde{W}}=M_2, \;m_{ \tilde{H}}=
|\m_{\eff}|,\  \mathrm{and} \;\varphi= \arg \left( \m_{\eff}\right)$)
the following weak basis interaction lagrangian 
arises:  
\bea\label{int:lagrangian}
-{\mathcal L}_{\mathrm{int}}= m_{ \tilde{W}}\overline{\tilde{W}}\tilde{W}
+ m_{ \tilde{H}} \overline{\tilde{H}} \tilde{H}
+\frac{g}{\sqrt{2}}(v_1 e^{-i \varphi} \overline{\tilde{W}}_R\tilde{H}_L+
v_2 e^{i \f_D} \overline{\tilde{W}}_L\tilde{H}_R+ \hc).
\eea
%
%
\section{Implications of indirect CP violation for the NMSSM}\label{scpv}
To explore the consequences  of SCPV on the upper bound of the lightest 
Higgs-boson mass we take into account
$\cp$ violation in $K^0$--$\bar{K}^0$ mixing. To accomplish this, 
we will compute the box-diagram 
contributions to $\epsi_K$ by applying the mass insertion
approximation. Let us start with the effective Hamiltonian 
governing $\D S=2$ transitions, 
which can be written as $\heff= \sum_i c_i {\mathcal O}_i$ . In the
presence of SUSY contributions the Wilson coefficients $c_i$ can be
decomposed as:
$c_i= c_i^W + c_i^{\higgs}+ c_i^{\chargino} + c_i^{\gluino} + 
c_i^{\neutralino}$. 
Given that the $\vckm$ matrix is real, 
and in the approximation of retaining only a 
single mass insertion in an internal squark line, we find that in the
present scenario with low $\tan\b$ we have a $c_i^{\chargino}$
dominance. Regarding the local operators 
${\mathcal O}_i$ \cite{fcnc:susy:constraints},  the $\D S=2$ 
transition is largely governed by the $V$--$A$ four-fermion operator 
${\mathcal O}_1 = \overline{d} \g^\m P_L s \overline{d} \g_\m P_L s$.
Therefore, we consider only the non-standard contributions to the
Wilson coefficient $c_1$, which are dominated by the diagrams depicted
in Figure~\ref{mia:leading}.
\EPSFIGURE{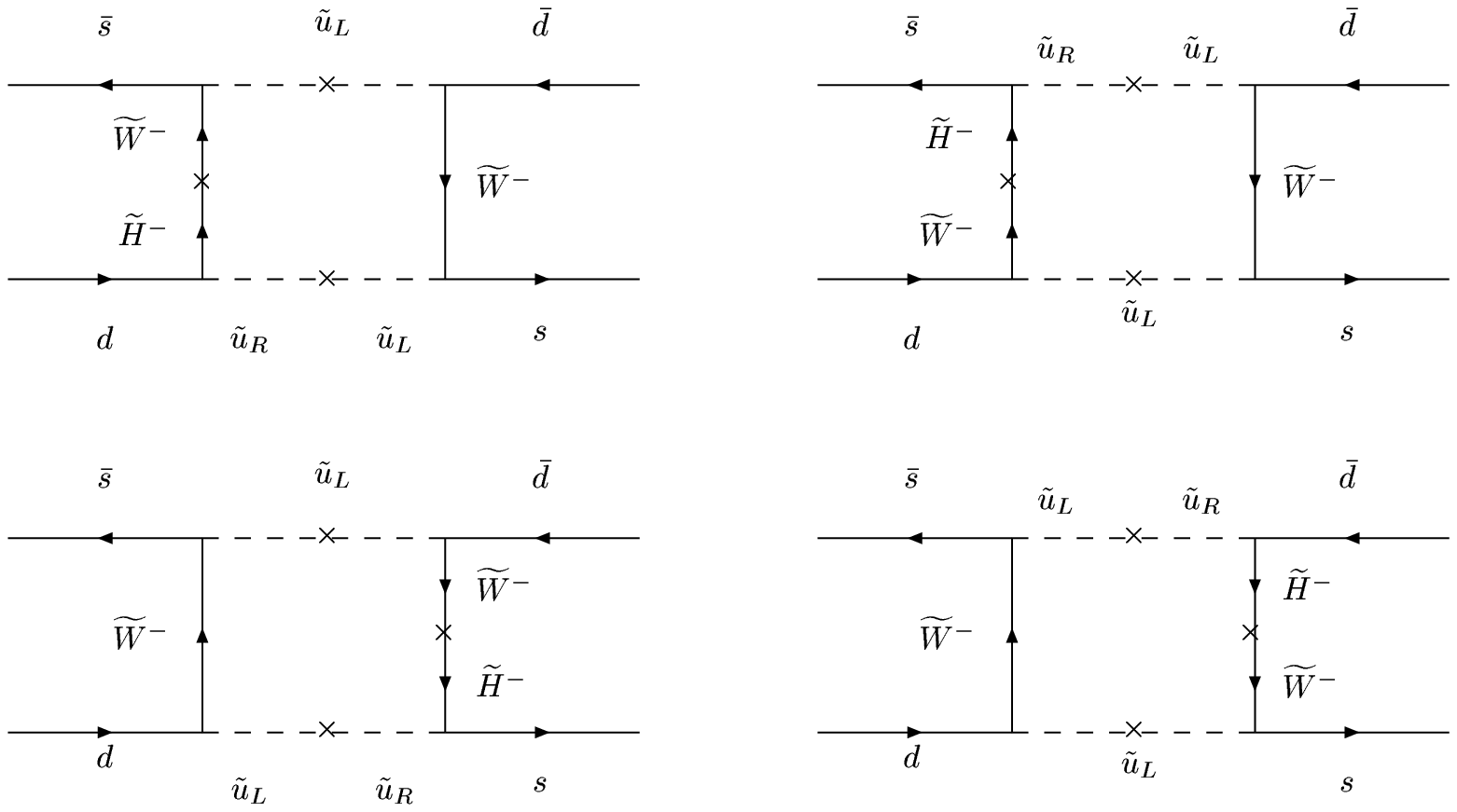,height=2.4in}
{The main contributions to $\epsi_K$ in the mass insertion approximation
with $W$-ino and Higgsino exchange.\label{mia:leading}}
In the limit of degenerate left-handed up-type squarks, keeping only
leading top-quark contributions and using the orthogonality of the $\vckm$,
we find that the imaginary part of the neutral
kaon mass matrix off-diagonal element is
\be\label{eK:eq:s4s13}
\Im \mathcal{M}_{12}=
\frac{2 G_F^2 f^2_K m_K m_W^4 }{3\pi^2\av{m_{\tilde{q}}}^8}
(V_{td}^*V_{ts}^{}) m_t^2 
\left|e^{i \phi_D}m_{{\wino}}+\cot\b m_{{\higgsino}}\right| 
\D A_U \sin(\varphi_{\chi}-\f_D)\;
{(M^2_{\tilde{Q}})}_{12}\;I_L \;.
\ee
In the above formula, $I_L$ is the loop function (see Ref.~\cite{SCPV}) and 
$\D A_U\equiv A_U^{13}-A_U^{23}$. From inspection of \eq{eK:eq:s4s13}, it is 
straightforward to conclude that in order to get a non-vanishing 
$\Im \mathcal{M}_{12}$ we need a theory of non-universal $A_U$ terms
(i.e. $\D A_U\neq 0$); in other words, it is not possible to saturate the 
observed $\cp$ violation in the $K$-meson system in the context of SUSY 
with a real CKM matrix and universal $A_U$ terms. 
Our results for the absolute value of 
$\epsi_K$ for various sets of SUSY parameters and low $\tan\b$
are reported in Table \ref{table:res} \footnote{For our numerical calculations,
we have used the nominal values
${(M^2_{\tilde{Q}})}_{12}/\av{m_{\tilde{q}}}^2=0.08 ,\; \\V_{ts}=-0.04,\;  
V_{td}=0.0066,\; m_t=175\,\GeV$ and 
$\D A_{U}=500\,\GeV$.}. \\
\TABLE{\small
\begin{tabular}{cccccccccc} \hline\hline
$|\epsi_K|$ &$\phi_D$  &$\phi_N$ & 
$m_{H^0}$ & $\av{m_{\tilde{q}}}$ &
$m_{\tilde{t}_R}$ &$\tan \beta$  &$\lambda$ & $v_3$\\
$(10^{-3})$&$(\mathrm{rad})$ & $(\mathrm{rad})$ &$(\GeV)$&$(\GeV)$&$(\GeV)$&
&&$(\GeV)$ \\ \hline 
$3.24$ &$4.71$   &$1.57$  &$99$ &$252$ &$235$ &$6.7$&$-0.03$ &$327 $ \\ 
$3.03$ &$0.89$   &$1.75$   &$97$ &$261$ &$168$ &$6.6$&$+0.33$ &$387$ \\ 
$2.75$ &$4.71$   &$4.71$   &$99$ &$232$ &$201$ &$9.2$&$-0.02$ &$221$ \\ 
$2.42$ &$1.96$   &$4.08$   &$94$ &$299$ &$174$ &$5.1$&$-0.06$ &$352$ \\ 
$2.10$ &$4.67$   &$4.75$   &$98$ &$279$ &$220$ &$7.8$&$+0.01$ &$142$ \\ 
$2.02$ &$4.68$   &$4.71$   &$92$ &$250$ &$152$ &$7.4$&$+0.02$ &$371$\\ 
$2.01$ &$4.18$   &$4.73$   &$96$ &$280$ &$232$ &$4.6$&$-0.01$ &$238$ \\
$1.31$ &$1.12$   &$4.72$   &$100 $&$273$&$241$ &$9.6$&$-0.01$ &$238$ \\ 
$1.29$ &$2.35$   &$4.70$   &$99$ &$258$&$230$ &$6.1$&$-0.13$ &$363$ \\ 
\hline\hline
\end{tabular}
\caption{Numerical values of $|\epsi_K|$ in the low $\tan\b$ region
for certain sets of model parameters that satisfy the minimisation 
condition of the Higgs potential.}\label{table:res}
}
From \eq{eK:eq:s4s13} it is clear that there is a linear dependence of $\epsi_K$ on the 
relative difference $\D  A_{U}$. In order to saturate the observed value of 
$|\epsi_K|$ \cite{PDG} and to obey present experimental 
limits on the sparticle spectrum, one has to take $\D A_{U}$ of order 
$500\,\GeV$. Values of $A_U^{i3}$ ($i=1,2$) around the $\TeV$ scale do not 
significantly affect the mass spectrum of the theory, and can 
account for values of the left-right mass insertions
$(\d_{LR}^U)_{i3}$  which are consistent with present experimental
bounds \cite{janusz}.\\
From Table \ref{table:res}, it is clear that we are in
the presence of large  $\cp$ phases, and hence potential problems with
the electric dipole moments (EDM's) of the electron and neutron.
Given the analytic results for the contributions to the EDM's of electron and 
neutron mediated by photino and gluino 
\cite{fcnc:susy:constraints}, together with the sets of parameters 
displayed in Table \ref{table:res} and the present experimental results of 
$d_n < 6.3\times 10^{-26}\, e\, \mathrm{cm}$ ($\cl{90}$) and
$d_e = 1.8\times 10^{-27}\, e\, \mathrm{cm}$ \cite{PDG},
the photino and gluino masses are 
required to satisfy  
$0.5 \, \TeV \lesssim m_{\tilde{\g}} \lesssim 2\, \TeV$ and
$2  \, \TeV \lesssim m_{\gluino} \lesssim 6\, \TeV$.
Such a hierarchy in the soft gaugino masses is
rather unnatural (since the masses of the squarks and $W$-ino 
are typically of the order $100$--$300\, \GeV$ in this
model). Moreover, masses of the superpartners of about $1\, \TeV$ 
may be in conflict with the cosmological relic density.
Finally, note
that the above-mentioned hierarchy for the spartners leads to an
unacceptable scenario for the lightest supersymmetric particle (LSP). In this 
case, the LSP would be either charged or would have a non-zero lepton number.
%
%
\section{Conclusions}
In this work, we have studied spontaneous $\cp$ violation in the 
context of the NMSSM, demonstrating that it is possible to generate sufficient 
$\cp$ violation in order to account for the magnitude of $\epsi_K$.
We have shown that the minimisation of the most general 
Higgs potential leads to an acceptable mass spectrum which is accompanied by
large $\cp$-violating phases. We have discussed that in order to 
account for  the  
observed $\cp$ violation in $K^0$--$\bar{K}^0$ mixing a rather 
low SUSY scale with 
$M_{\susy}\approx 300\,\GeV$ (i.e.~light squark and $W$-ino masses) 
and a non-trivial flavour structure of the soft SUSY-breaking trilinear 
couplings $A_U^{i3}$ ($i=1,2$) are required. As a consequence, the
parameter space is severely constrained and the mass of 
the lightest Higgs boson is further diminished,
and it turns out to be no greater than $\sim 100\,\GeV$ for the case 
of low $\tan\b$ ($\lesssim 10$).
We have also argued that it may be difficult to reconcile the large-phase solution
with the severe constraints on the EDM's of electron and neutron.
Therefore, the implications of the 
EDM bounds on the parameter space will be a great challenge for SUSY
models with spontaneous CP violation.
%
%

%
\end{document}